Title: Comment on Suzuki's rebuttal of Batra and Casas

By Yoshiaki Nakada, Division of Natural Resource Economics, Graduate school of Agriculture, Kyoto University, Kita-Shirakawa-Oiwake-cho, Sakyo, Kyoto, Japan

E-Mail: nakada@kais.kyoto-u.ac.jp,

Abstract: Batra and Casas (1976) claimed that 'a strong Rybczynski result' arises in the three-factor two-good general equilibrium trade model. In subsequent comments, Suzuki (1983) contended that this could not be the case. Among his comments, Suzuki found that the set of three equations holds for the Allen-partial elasticity of substitution under the assumption of perfect complementarity, and he applied these to his analysis. In the following, I demonstrate that these are impossible, hence his dissenting proof is not plausible.

1. Introduction

In a rebuttal to the article by Batra and Casas (1976) on functional relations in a three-factor two-good neoclassical model, Suzuki (1983) provided a dissenting proof. In a review of this rebuttal, I find that Suzuki made an error in his proof which renders it implausible. Batra and Casas (1976) were looking at the relation between outputs and factor endowments, which can be found in Theorem 6 (p34). According to Suzuki, they contended that 'if commodity 1 is relatively capital intensive and commodity 2 is relatively labor intensive, an increase in the supply of labor increases the output of commodity 2 and reduces the output of commodity 1.'

In presenting his proof, Suzuki argues that, 'their conclusion is not true […] The argument is developed in terms of their notations in a model in which capital and land are perfectly complementary to each other in the production of each commodity.' Suzuki found that the set of three equations holds for the Allen-partial elasticity of substitution (hereinafter, AES) under the assumption of perfect complementarity. On this, see eq. (1) shown below. In applying this to his analysis, he tried to show that, in contrast to Batra and Casas, 'a strong Rybczynski result' (to use Thompson's terminology, 1985, p617) does not hold in case of perfect complementarity.

This rebuttal was subsequently confirmed by Jones and Easton (1983, p67) and Thompson (1985). In a summary of the article, Thompson (1985, p617) suggests that,

'Batra and Casas (1976) claim a strong Rybczynski result [which holds for a two-factor, two-good model] stated in terms of extreme factors is also found in the three-factor model. Suzuki (1983) and Jones and Easton (1983) point out, as is done here, that a strong Rybczynski result is not necessary.' In this paper, I return to Suzuki's original rebuttal, finding that elements of his disproof are questionable.

2. Discussion
   Suzuki (1983, p142) stated,

'Suppose that land and capital are used in a fixed proportion (bj) for the relevant w/r and w/t ratios in each industry. That is,

$$C_{Kj} = b_j C_{Tj}, \; j = 1, 2, […]$$

where bj are constants.'

Here, w is the wage rate, r the rental on capital, t the rent of land; $C_{ij}$ is the requirement of the ith input per unit of the jth good; T is land, K capital, L labor. Suzuki continued,

'Under the assumption of the complete complement relation between land and capital,

$$C_{Kj}^* = C_{Tj}^*,$$

or

$$\sigma^j_{KK} = \sigma^j_{KT} < 0; \; \sigma^j_{LK} = \sigma^j_{LT} > 0, \text{ and } \sigma^j_{KT} = \sigma^j_{TT} < 0.', \quad (1)$$

where the asterisk denotes the rate of change (e.g. $C_{Kj}^* = d\,C_{Kj} / C_{Kj}$); $\sigma^j_{ik}$ is the AES between the ith and the kth factors in the jth industry. For additional definition of these symbols, see Batra and Casas' original (1976, pp22-24).

In the final test, Suzuki substituted eq. (1) into eq. (26) found in Batra and Casas (1976, p32). In doing so, he found that $X_1^*/L^*$ is positive, and thereby concluded that their result was not true [see Suzuki (1983, p143)]. $X_j$ is the amount produced of the jth good (j=1, 2), and L is the supply of labor [see Batra and Casas (1976, p22)].

In summary, Suzuki found that eq. (1) holds for the AES under the assumption of perfect complementarity, and used this in his disproof. However, Batra and Casas (1976, p33) derived the relationship for AES on the assumption that the production functions were strictly quasi-concave and linearly homogeneous, i.e.:

$$\sigma^j_{KK}\sigma^j_{TT} - (\sigma^j_{KT})^2 > 0. \quad (2)$$

If we compare this inequality (2) with eq. (1), we find that the latter is not consistent with the former. Hence, the Suzuki's result is impossible. Specifically, Suzuki failed to explain what perfect complementarity implies. In sum, his proof is not plausible.

3. Conclusion

In their original, Batra and Casas (1976) claim that 'a strong Rybczynski result' is found in the three-factor two-good model. Suzuki dissented, using the fact that eq. (1) holds for AES under the assumption of perfect complementarity, in his disproof. This element, I find, is impossible, which renders the proof implausible.

Suzuki failed to notice the important relationship among AES as shown in eq. (2) which Batra and Casas (1976) derived from the assumptions for the production functions. Suzuki's fellow researchers, including Jones and Easton (1983) and Thompson (1985), did not catch this oversight. It is possible that, due to the large number of formulae found in the article by Batra and Casas, readers simply failed to catch important errors. Regardless of the cause of this error, we are now left to search for a new sufficient condition for 'a strong Rybczynski result' to hold (or not to hold).

Acknowledgement: I would like to thank Hart Feuer, Kyoto University for giving many helpful comments on language, and thank Raveendra N. Batra, Southern Methodist University and Thompson Henry, Auburn University for reading an earlier version and agreeing my conclusion.